\documentstyle[onecolumn,psfig]{mn}

\title[Baryon Mass Function]
{\bf The Baryonic Mass Function of Spiral Galaxies: Clues to
Galaxy Formation}
\author[Paolo Salucci and Massimo Persic]
{Paolo Salucci$^{1}$ and Massimo Persic$^{2}$ \\
$^1$ SISSA -- International School for Advanced Studies,
via Beirut 2-4, I--34013 Trieste, Italy \\
$^2$ Osservatorio Astronomico, via G.B. Tiepolo 11, I-34131
Trieste, Italy \\
$ ~~$ {\tt salucci@galileo.sissa.it, persic@ts.astro.it} }

\begin{document}

\maketitle
\def\omes{\Omega^{^{_{\rm S}}}_b} 
\def\omel{\Omega_{Ly\alpha}}
\def\mly{M_{Ly\alpha}}
\def\mincir{\raise
-2.truept\hbox{\rlap{\hbox{$\sim$}}\raise5.truept
\hbox{$<$}\ }}
\def  \magcir{\raise
-2.truept\hbox{\rlap{\hbox{$\sim$}}\raise5.truept
\hbox{$>$}\ }} 
\def\rn{$ ~{0.06 \, h_{50}^{-2} \over \Omega_{BBN} }~$} 
\def\ref{\par\noindent\hangindent 20pt}

\begin{abstract}
We compute the baryonic mass function, $\psi^{^{_{\rm S}}}(M_b) d{\rm log} 
\,M_b$, of disc galaxies using the luminosity functions and baryonic 
mass-to-light ratios reliable for this goal. On scales from $10^8 M_\odot$ 
to $10^{11} M_\odot$ this function is featureless, $\psi^{^{_{\rm S}}} 
\propto M_b^{-1/2}$. Outside this mass range $\psi^{^{_{\rm S}}}$ is a strong
inverse function of $M_b$. The contributions to the baryon density $\omes$
from objects of different mass indicate that spirals have a characteristic
mass scale at $M_b^{\oplus} \simeq 2 \times 10^{11} M_\odot$, around which
more than 50\% of the total baryonic mass is concentrated. The integral value, 
$\Omega_b^{^{_{\rm S}}}=(1.4 \pm 0.2) \times 10^{-3}$, confirms, to a higher 
accuracy, previous evidence (Persic \& Salucci 1992) that the fraction of BBN 
baryons locked in disc galaxies is negligible and matches that of high-$z$ 
Damped Ly$\alpha$ systems (DLAs). We investigate the scenario where DLAs 
are the progenitors of present-day spirals, and find a simple relationship 
between their masses and HI column densities by which the DLA mass function 
closely matches that of spiral discs.
\end{abstract}

\section{Introduction}

The visible mass of a galaxy is intimately related to the
process of its formation, indicating how, during its cosmological history,
the primordial baryon-to-total density ratio $\Omega_{BBN}/\Omega$ has
been modified. Furthermore, the inventory of stars and gas in galaxies
sheds light on where the BBN baryons reside at the present time, and 
helps to investigate the nature of the baryonic structures observed at
high $z$. 

In this connection, a step forward was made by Persic \& Salucci (1992; 
hereafter PS92) who estimated the cosmological density of the baryonic matter in
spirals $\Omega_b^{^{_{\rm S}}}$ (and in other bound systems) by averaging
the spirals' disc masses $M_\star(L_B)$ over their luminosity function
(LF), $\phi^{^{_{\rm S}}}(L_B) dL_B$: 
$$ 
\omes ~=~ {1\over{\rho_c}} ~
\int^{L_B^{max}}_{L_B^{min}} M_\star(L_B) ~ \phi^{^{_{\rm
S}}}(L_B) ~ dL_B
\eqno(1)  
$$ 
where  $\rho_c$ is  the
critical mass density of the Universe \footnote{$\rho_c \equiv {3 \, H_0^2
\over 8 \pi G }$. We use $H_0=75$ km s$^{-1}$ Mpc$^{-1}$ throughout. No result 
in this paper depends crucially on the value of $H_0$.}, 
and $L_B^{max}$, $L_B^{min}$ have their obvious meanings. By
means of the disc mass vs. luminosity relationship [derived by applying
the Persic \& Salucci (1990a) method of  mass decomposition to 
$\sim 60$ rotation curves] and adopting the Schechter LF (derived from the
AARS survey, $\sim$ 200 objects; Efstathiou et al. 1988),  PS92 found
$\Omega_b^{^{_{\rm S}}}=7^{+6 }_{-4}\times 10^{-4}$ implying that only few 
percent of the cosmologically synthesized baryons are detected today in
spirals (or in other bound structures, PS92). 
 
Our knowledge of the properties of spirals has improved enormously in the
past few years. In fact, with respect to PS92, we can now rely on:  {\it
(a)} LFs deeper by about 5 magnitudes and resolved by  Hubble Type; {\it
(b)}  HI mass estimates for objects   spanning
the whole luminosity range of disc systems; {\it (c)} 
determinations of disc masses for Sa and dwarf spirals; {\it (d)}
refined  derivations of disc and bulge  masses  in  normal spirals.

Based on this new knowledge, we investigate two crucial (and
timely) cosmological issues: the baryonic mass function of disc systems,
and the connection between the population of local spirals and that of
high-$z$ Damped Ly-$\alpha$ clouds (DLAs). 

As the halo mass function carries direct information on the spectrum of
primordial cosmological perturbations and the luminosity function reflects
the evolution of the stellar populations of galaxies, the baryonic mass
function of disc systems (DMF) is a unique probe into the late stages of
their formation, including the dark-to-luminous coupling that has formed
spirals as we see them today.
 
The plan of this paper is as follows. In section 2 we introduce
the spiral LF and (disc mass)-vs.-luminosity relation. In sections 3
and 4 we estimate the DMF and the amount of baryons locked in spirals. In
section 5 we investigate the connection between spirals and DLAs. 
Finally, in section 6 we briefly discuss some cosmological implications.

\section{Basic properties of spirals}

\subsection{Luminosity function}

Recent large surveys have obtained the LF of galaxies down to M$_B \sim
-12.5$ \footnote{In this paper the magnitudes are in the $B_{ESO}$ system.
When transformed from the $B_J$ system, the adopted conversion is
$B_{ESO}= B_J+ 0.2$.}. The standard Schechter (1976) function,
$$ 
\phi(L_B) ~ dL_B  ~=~ \phi_0 ~ \biggl({L_B \over 
L_B^*}\biggr)^{-\alpha }~ e^{-{L_B \over  L_B^*}} ~ {dL_B \over L_B^*}\,, 
\eqno(2a)
$$ 
is generally found to give an excellent fit to the data (Efstathiou et al. 
1988; Loveday et al. 1992; Marzke et al. 1994; Lin et al. 1996; Radcliffe 
et al. 1998). However, for M$_B \magcir -15$, there is a significant 
excess of objects above the Schechter prediction (e.g., Loveday 1998). We
consider the two cases seperately. 
\medskip

 
\begin{figure}
\vspace{6.7cm}
\includegraphics{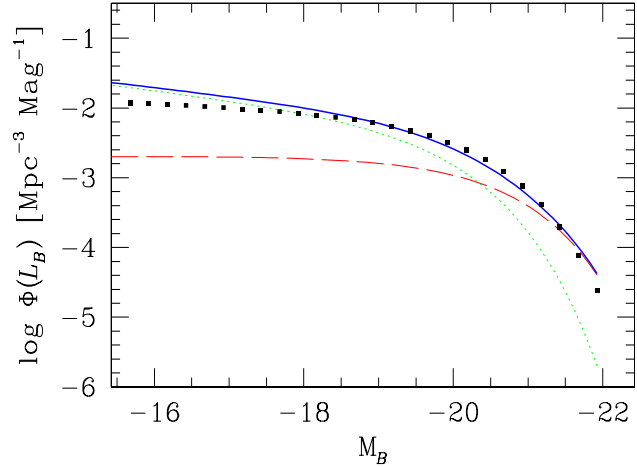}
\caption{
The luminosity function of spiral galaxies  for M$_B<-15$. The 
spectral Sa and Sb-Im luminosity functions  are indicated by a dashed and
a dotted line, respectively: 
the resulting total spiral LF is indicated by a solid line. The filled 
squares represent   the morphological spiral LF of Marzke et al.
(1998).}
\end{figure}

\begin{figure}
 
\vspace{7.5cm}
\includegraphics{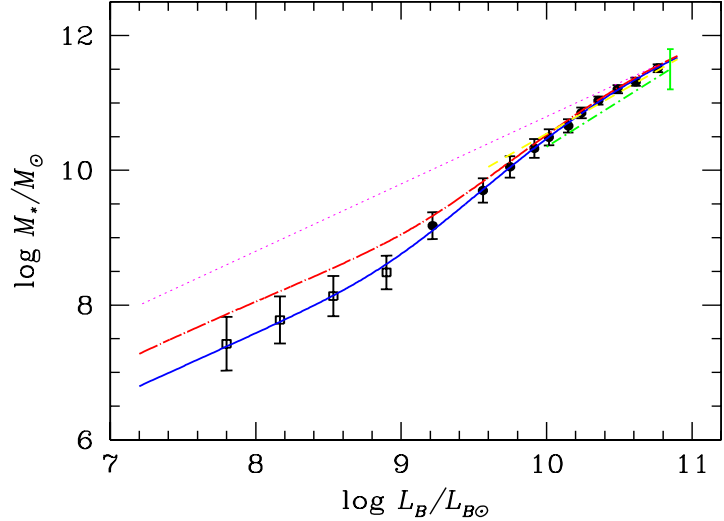}
\caption{
The stellar $M_\star(L_B)$ and baryonic $M_b(L_B)$ relations (solid
and dot-dashed lines). Filled circles 
and empty squares represent the data for late spirals and dwarf irregulars. 
The short-dashed line and the dot-short-dashed  line denote $M_\star (L_B)$
for, respectively, maximum-disc and photometric masses. 
We also show $M_b(L_B)$ corresponding to a  
(baryonic mass)-to-light ratio assumed to be constant with luminosity
(dotted line). 
}.

\end{figure}

\noindent
{\it (i):} M$_B < -15$.
\smallskip

\noindent 

With respect to the earlier AARS survey, the Autofib Redshift Survey
(1700 objects; Ellis et al.  1996) has filled the gap $B=17-21$ in the coverage
of apparent magnitudes, and has significantly increased the size of the sample
down to $B=22$.  This has made it possible to determine the LF for each spiral
Hubble
Type, down to M$_B \simeq -15.5$, with an accurate estimate of the faint-end
slope and the overall normalization (Heyl et al. 1997).  In detail, we take the
Heyl et al.  (1997) Sa LF and we combine the remaining three Sb-Sdm LFs into a
single late-type LF (see Fig.1). The parameters of these two Schechter functions are
reported in Table 1. The values of $\alpha$, $\phi_0$, and $M_B^*$
vary, field by field and survey by survey. The typical uncertainties
involved in the determination of $\alpha$, $\phi_{0}$ and $M_{B}$
(Lin et al. 1996; Radcliffe
et al. 1998) are : 0.05, $20\%$, and 0.1 mag, respectively.

The observational input is further increased  
by considering also the spiral LF of Marzke et al. (1998),   built  
from the morphology of the objects rather than from their spectral features:
the relative  Schechter parameters are given in Table 1. For later
considerations let us notice that, from the  above LFs, the value of the
parameter $L_B^*$ relative to the entire  spiral population is:
$L_B^*\simeq 2\times 10^{10} L_{B\odot}$.   

Finally, for $M_B < -15$, both LFs neglect low-surface-brightness (LSB) galaxies, 
which constitute a very small (and uncertain) fraction of the whole population of 
disc systems at these luminosities (Sprayberry et al. 1997; see also next subsection).
\smallskip

\noindent

{\it (ii)} M$_B > -15$.
\smallskip

\noindent
The 2.4-million-galaxy APM survey has provided the first direct
information on 
the very faint end of the LF, reaching M$_B \simeq -12.5$. 
It has found, faintwards
 of M$_B \sim -15$,  an increasing excess of
galaxies with respect to the Schechter profile, mostly of
late-type
morphology and low surface brigthness (Loveday 1998; see also
Marzke et
al. 1994, Lin et al. 1996, Zucca et al.  1997). This rapid increase
can be
taken into account by adding to the standard LF an additional
power-law
term that switches on at M$_B \magcir -15$ and becomes
comparable to the
Schechter term at $M_B^t \sim -14$ (see Loveday 1998). 

Thus, the luminosity function of disc systems can be written as: 
$$
\phi^{^{_{\rm S}}}(L_B) ~=~ \phi (L_B) + \phi(L_B^t)\biggl( {L_B
\over 
L_B^t} \biggr)^{-2.7}\,,
\eqno(2b)
$$
with $\phi(L_B)$ given by eq.(2a) and $L_B^t \simeq 8 \times
10^7 L_\odot$ 
(Loveday 1998).

\subsection{Disc masses} 

\subsubsection{Sb--Im galaxies}

Using a very large dataset of 1100 high-quality rotation curves (RCs)
and a newly devised 
method of mass modelling, Persic et al. (1996) have
determined the stellar 
disc and bulge masses of Sb-Sdm normal spirals 
($-18 > {\rm M}_B > -22.5$) as a function of blue luminosity
(see also Salucci \& Persic 
1997). In addition, recent RC analyses have
provided the stellar 
disc masses for a number of dwarf galaxies, $-15.5 \magcir
$M$_B \magcir -18$ [see Salucci \& Persic (1997)
and references therein]. Combining these results we find ($L_{10} \equiv 
10^{10}\ L_{B\odot}$): 
$$
M_\star(L_B) ~=~ 3.7 \times 10^{10}   
\biggl[ \biggl({L_B \over L_{10}}\biggr)^{1.23} g(L_B) ~+~ 
9.5 \times 10^{-2} ~ \biggl({L_B \over L_{10}}\biggr)^{0.98} \biggr]\  M_\odot
\eqno(3a)
$$
$$
g(L_B) ~=~ exp\biggl[-0.87 ~\times ~ \biggl({\rm log}{L_B \over
L_{10}}  -0.64\biggr)^2 \biggr]\,,
\eqno(3b)
$$
that links the stellar mass and the luminosity of disc systems in a way
quite different from the simple $M_b \propto L_B$ law (see Fig.2).

The  uncertainty ${\delta M_\star \over M_\star}\big|_{th}$  ranges 
from $5\%$ at the 
highest luminosities to $30\%$ at  the 
lowest luminosities. However,  the actual disc mass variance 
for galaxies of a given luminosity,   ${\delta M_\star
\over M_\star}$, is larger in that it also includes a variance due to    
observational errors plus an intrinsic component 
${\delta M_\star \over M_\star}\big|_{cosm}$. This latter uncertainty, which
    must be
taken into account when   convolving     the mass-to-light
ratios with the luminosity function,   can be  estimated
from the values of the disc masses    derived  for      individual galaxies.
 Using  $\sim 100$ objects   in Persic \& Salucci
(1990a), Salucci et al. (1991), and Persic et al. (1996), we conservatively estimate: 
 $\delta {\rm log} M_\star^{cosm} < 2.3 \,
{\delta M_\star\over M_\star}  \simeq 0.2 \ dex $. 
 
The quantity $\delta {\rm log} M_\star^{cosm}$ is often related to
 the  mean scatter of the Tully-Fisher (TF) relation,  
$\sigma_{{\rm log}V}\simeq 0.10 \ dex $ (e.g.,  Willick 1998).   
We stress that  there is no {\it direct} link 
 between these two quantities.   In fact, 
 at a fixed luminosity:  $ 2 \ \sigma_{{\rm log}V} \simeq  \Sigma(M_\star) + 
\Sigma(\beta) + \Sigma(L/R_{opt})$ (Salucci et al. 1993),
 where  
$R_{opt}$
 is the disc size, $\beta$ is the disc-to-total mass fraction
 inside  $R_{opt}$, 
 and $\Sigma(x)$ is    the
     cosmic log-variance 
    of the quantity ${\rm log}x$ convolved with its measurement errors. 
    Nevertheless,   
  since   $\Sigma(L/R_{opt}) = 0.05- 0.1 $ and  
$ \Sigma(\beta) \sim 0.05$ (e.g., Persic et al. 1996), we can state 
that   the variance of the   TF relation and that of eq.(3) are
mutually consistent.
  
Note that the mass vs.  luminosity relation 
assumed here is robust with respect to the
particular method employed in   deriving  the disc masses. In Fig.2, we  plot the 
 $M_\star$--$L_B$ 
relationship corresponding to $M_\star(L_B)$ obtained {\it (1)} as maximum-disc 
solutions (Persic \& Salucci 1990b), and {\it (2)} via 
stellar population synthesis models (Salucci et al. 1991). In both cases, 
  the  
differences with the RC-slope-best-fit values used  to derive   eq.(3) are 
irrelevant to the main results  of this paper.

\subsubsection{Sa galaxies}

The bulge and disc  masses of  Sa galaxies are  obtained 
from:  {\it (a)}  dynamical modelling of the line-of-sight  
dispersion velocity  profiles (of either the stars and/or the gas)
and/or of the extended HI and optical RCs; and {\it (b)}  
comparing the observed 
galaxy  spectra with  
stellar population synthesis templates (e.g.: Corsini et al. 1999; Bertola et al. 
1993, 1998; Silva et al. 1998; Honma \& Sofue 1997; Quillen \& Frogel 1997; 
Jablonka \& Arimoto 1992). Given that Sa galaxies contribute 
significantly to the spiral LF 
only in the  very small range of magnitudes  $-22<{\rm M}_B <-21$, 
the $\sim$20 objects available in the literature turn out to be 
sufficient to determine an average
mass-to-light ratio in this region   and
 to  establish its 
 trend with  luminosity. 
 We find that the stellar
 masses increase with luminosity
slightly more steeply  than linear, in good agreement  with eq.(3). 
However, the  mass-to-light ratios of Sa galaxies are  
a factor of order two larger than the corresponding values for late spirals.    
We therefore  assume:
$M_\star^{\rm Sa}(L_B) ~=~ 1.5 \times M_\star^{\rm Sb-Sdm}(L_B)$. 
[The main results of this paper do not depend on the (suitable) actual value 
assumed for this proportionality constant.]
   
\begin{figure}
\vspace{6.7cm}
\includegraphics{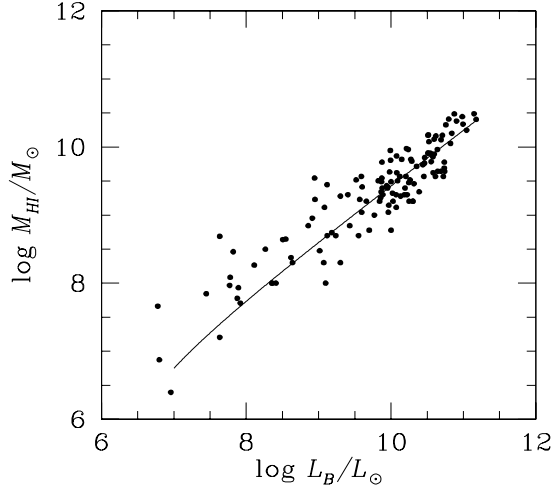}
\caption{ The HI mass vs. $L_B$ relation for disc galaxies. 
The data are from Hoffman et al. (1996: Table 2) and Rhee
(1996: Table 7.1). 
} 
\end{figure}

\subsubsection{Gas content}

HI masses have recently been derived, through HI-flux
observations, for
two large samples of spirals that include also irregular dwarfs
(Hoffman
et al. 1996; Rhee 1996). The data for the combined sample
($10^7 \ L_\odot \mincir
L_B \mincir 10^{11}L_\odot$) imply a strong $M_{HI}(L_B)$
relation (see
Fig.3), especially so if it is reckoned that this is steeper at lower
masses
than at higher masses (Hoffman et al. 1996; Salpeter \&
Hoffman 1996). In
detail, we have: 
$$
{M_{HI} \over M_\odot} ~=~ 1.6 \times 10^6 ~\biggl( {L_B \over
10^6
     L_\odot}\biggr)^{0.81}
     \biggl[1-0.18\, \biggl( {L_B \over 10^8 L_\odot} \biggr)^{-0.4}
     \biggr] 
\eqno(4)
$$
(see Salpeter \& Hoffman 1996 for further details). To take the
helium
contribution into account, we multiply the r.h.s. of eq.(4) by 1.33. 

Two other disc-like gaseous components are present in spirals:
{\it (i)}
molecular gas ($H_2$, CO) which, however, being distributed as
the exponential stellar disc, is already taken into account by
$M_\star$; and
{\it (ii)} ionized hydrogen (found by pioneering
H$\alpha$-emission
measurements, see Bland-Hawthorn et al. 1997), the mass of which is,
however, 
difficult to estimate, also because the HI/HII transition has a
sharp edge
(Corbelli \& Schneider 1997). For the time being we consider the
HII component
as provisionally unidentified baryonic dark matter.
\bigskip

The baryonic masses of disc systems are then obtained from
eqs.(3) and (4): 
$$
M_b =1.33\, M_{HI}+M_\star
\eqno(5)
$$ 
We emphasize that these determinations  are more
accurate than 
those in PS92 in the following respects: {\it (i)} they  hold down to $\sim$4
magnitudes 
fainter; {\it (ii)} they include the 
gaseous disc mass; {\it (iii)} they 
bear
smaller  theoretical and observational uncertanties; and {\it (iv) } they allow
us to take into account    the population of 
 LSB galaxies whose 
 HI masses,  much larger than 
their corresponding  stellar masses (e.g., de Blok et al.
1996), 
 are  well represented
by  eq.(4).
It is  worth noticing that these  improvements, while absolutely needed
 to  derive the 
spiral baryonic mass function, imply refinements on   
$\omes$ (as estimated by  PS92) that are too small to affect  the PS92
 claim      
 $\omes<< \Omega_{BBN}$.

\section{\bf The disc mass function}

We investigate disc systems in the luminosity range between $
\simeq 10^7
L_\odot$ and $L^{max}_B \simeq 8 \times 10^{10} L_\odot$,
corresponding to
the mass range between $M_b(10^7L_{B\odot})\simeq 10^7 M_\odot$ and
$M_b^{max}  \equiv  M_b(L_B^{max})\simeq 4
\times 10^{11} M_\odot$.  Massive ($M_b \magcir 10^{11} M_\odot$)
discs are rare objects, while light discs 
($10^7M_\odot \mincir M_b \mincir 10^9 M_\odot$)
constitute the most
numerous population of galaxies in the Universe: the lowest
masses considered
here reflect the lack of suitable data, not the lack of objects.
Stellar
disc masses are known only for $L_B \magcir 5 \times 10^7
L_\odot$:
however, at these luminosities the gas mass given by eq.(4) is,
by far,
the main baryonic component. 

By using  eqs.(2)-(5) we derive the  disc (systems)  mass function
(DMF),
 $ \psi^{{_{\rm S}}}(M_b)$, defined  as\footnote{ with $\int_{range\  L_B}
\phi
(L_B) dL_B = \int_{range\  M_b}
\psi(M_b)\,  dM_b$ from  the conservation of the number
of galaxies.}: 
$$
\psi^{{_{\rm S}}}(M_b)~ d{\rm log}\, M_b ~=~ \phi^{^{_{\rm
S}}}(L_B(M_b)) ~ {dL_B \over dM_b} ~  d{\rm log}\, M_b \,.
\eqno(6)
$$
We  take  into account   the    scatter of the $M_b(L_B)$ 
relation by convolving the r.h.s of eq.(6) with a  Gaussian of
half-width $\delta {\rm log} M_b=0.2$ corresponding  to the maximum 
cosmic variance in relationship (3).

\begin{figure}
\vspace{7.2cm}
\includegraphics{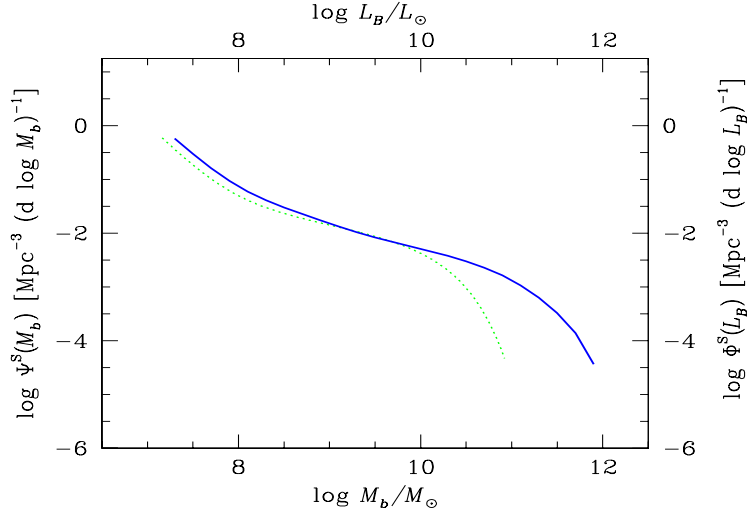}
\caption{
The luminosity function (dashed line) and the baryonic mass
function
(solid line) of spiral galaxies.  
}
\end{figure}

Fig.4 shows that the DMF can be well described by a  power law
$\psi^{^{_{\rm
S}}} 
\propto M_b^{-1/2}$ over three decades in mass, different
from the LF that 
can hardly be reproduced by a power law over more than a
decade in luminosity. 
A good fit for the   DMF is:
$$ 
\psi^{^{_{\rm S}}} =1.2 \times  10^{-3} \Big[1+ \Big({M_b\over M_t}\Big)^{-1.46}
\Big] 
\Big({M_b\over M_b^{\oplus}}\Big)^{-0.46} e^{-\big( {M_b\over
M_b^\oplus}\big)}\,,
\eqno(7)
$$
where $M_b^{\oplus}=2.7 \times 10^{11}  \ M_\odot$ and 
$M_t= 6.7 \times  10^7   M_\odot$.

At the 
highest masses, the DMF shows a sharp cutoff that noticeably 
occurs at a mass 
$\simeq M_b^{\oplus}>3 \ M_b(L_B^*)$
(with $M_b(L_B) \simeq 8 \times 10^{10} M_\odot$), quite
different from the baryonic mass corresponding to 
 $L_B^* \simeq 2 \times 10^{10}$, 
the 'knee" of the spiral LF (e.g.: Heyl et al. 1997; Marzke, 1998).
 This supports the scenario in which
spirals were formed under 
an upper mass limit, $\simeq M_b^{max}$, due to the inability of
a larger 
baryonic mass to cool fast enough to settle into a disc within a
Hubble time 
(Rees \& Ostriker 1977; see also Thoul \& Weinberg 1996). This
sharp cutoff 
gets broadened in the LF (Figs. 4 and 5) 
because the 
radiating efficiencies of stellar discs of a given mass have a
significant 
scatter, due to differences in the discs' stellar populations (see Oliva
et al. 1995).
 
At $L_B<10^8 L_\odot$, there is a steepening of the DMF,
parallel to that
of the LF, that might hint to a bimodal distribution. In any case,
also in
the DMF there is no sign that downwards of the smallest
observed masses,
$M_b \sim 10^7\ M_\odot$, the objects are disappearing. 

Let us notice that the intrinsic variance  of the $M_b(L_B)$ 
relationship has  negligible effect on the  DMF, both
because the latter is essentially 
a power law and because   $\delta {\rm log} M_b$
is much smaller than the  range of  log$M_b$.

\begin{figure}
\vspace{6.3cm}
\includegraphics{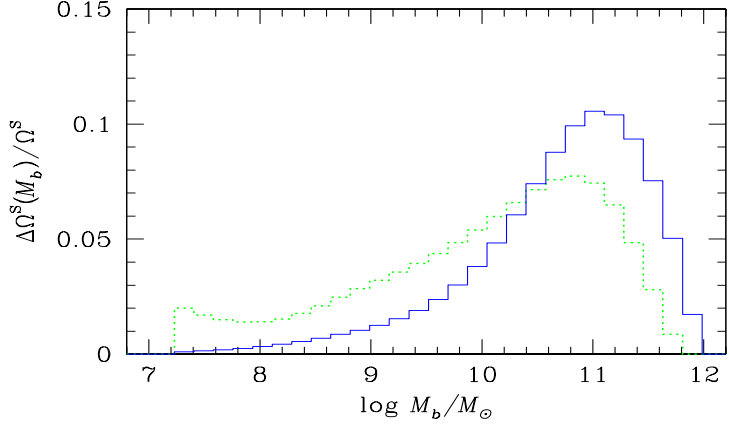}
\caption{
The contribution of spirals  of different mass
to 
the baryonic mass/luminosity density (solid/dotted line) of the
Universe. 
}
\end{figure} 

Finally,   the main features of the DMF do not
depend 
on which (suitable) LF  is assumed:    the
Marzke et al. (1998) LF yields a DMF 
very similar to eq. (7)  (see Fig.7 in the Appendix.)   

Despite the baryonic mass function  being essentially
featureless, spirals do have a
characteristic mass scale: $M_b^{\oplus}$. In  Fig.(5) we show
$\Delta
\Omega_b^{^{_{\rm S}}}(M_b)$, the differential contribution to
$\omes$
from spirals belonging to intervals of width 0.3 centered on
log$M_b$:
in spite of the wide range of baryonic mass (5 decades: 24
intervals), the
major contribution to $\omes$ ($\magcir 60 \%$) comes from
objects within
a factor 3 of $M_b^{\oplus}$. Thus, spirals 
do form with masses in
the range
$10^7M_\odot - 10^{12}\ M_\odot$: but most of the galactic
baryons are
actually locked at one specific mass scale, $M_b^{\oplus}$,
higher than
the mass of an $L_B^*$ galaxy and much higher than the mass of the
great
majority of the objects. Actually, more than $75 \%$ of the total
baryonic
mass is segregated in less than $0.5\%$ of the objects! 
Conversely, the
light is much more evenly distributed among galaxies of
different
luminosities:  e.g., objects with $M_b < 10^{-2} M_b^{\oplus}$
still
contain more than $20\%$ of the total light. 

As a consequence, the "typical" scale $L_B^*$ does not have a
cosmological
significance, neither as the most relevant mass scale nor as the
mass at
which the mass function cuts off. At $L_B^*$ the mass function
is still
increasing, with most of the cosmological baryon mass density
located at
$L_B > L_B^*$.

\section{ The baryon content of spirals}

Let us refine the PS92 estimate of the cosmological density of
the stellar
component in disc systems with the improved LF and
mass-vs.-light relation
of the previous sections. From eqs.(2)-(5) we have: 
$$ 
\Omega_{b}^{^{_{\rm S}}} ~=~ 1.44^{+0.15}_{-0.20} \times
10^{-3}\,, 
\eqno(8) 
$$ 
marginally larger and substantially more accurate than the PS92
estimate. The Marzke et al. (1998) LF yields:
$\omes = 
(1.3 \pm 0.25) \times 10^{-3}$.
Let us notice that if we  neglect the (relevant) presence of dark
matter
inside $R_{opt}$  and/or the luminosity dependence of the
(stellar +
gas) mass-to-light ratio, we are led to overestimate $\omes$ by
up to a
factor of 4 and to fictitiously reduce the observed discrepancy
between
$\Omega_b^{^{_{\rm S}}}$ and $\Omega_{BBN}$.

As a finer detail in eq.(8), the amount of HI+HeI in spirals is
$$
\Omega_g({\rm S}) ~=~ (1.7 \pm 0.8) \times 10^{-4}\,.
\eqno(9) 
$$ 
This is in good agreement with the cosmological HI content,
$\Omega_{gas}=
(3.3\pm 1) \times 10^{-4}$, estimated by integrating the HI
mass function
of Zwaan et al. (1997), especially considering that the latter may
include
the S0 discs, neglected in our present estimate.

With the caveat that the masses of ellipticals and spheroidals
are still
uncertain by $40 \%$ (Salucci \& Persic 1997), let us estimate $\Omega_b^{\rm
E}$, the
cosmological baryonic mass density in ellipticals and S0. By
means of the
E/S0 Autofib LF and the PS92 mass--luminosity relation (see
also Salucci \& Persic 1997), we
find $\Omega_B^{\rm E} \sim (2 \pm 1) \times 10^{-3}$, in
agreement with
PS92 and with Salucci et al. (1999). Spheroids then store about
 twice as much baryonic
matter as discs,
differently from the common belief of an equipartion of the
baryonic
matter between discs and spheroids (e.g., Schechter \& Dressler
1987).
Instead, we point out that  with respect
to
spirals, ellipticals have about 1/10 of the number density, 1/2
of the
luminosity density and 2 times the baryonic mass density.

The total amount of baryons in galaxies is then $(3 \pm 1)
\times
10^{-3}\rho_c$. It is remarkable that the star formation rate
density,
derived from the [OII] luminosity function (Hogg et al. 1998),
when
integrated over the lifetime of the universe, implies  [caveat  
a set of (well justified)  assumptions on  the IMF, metallicity and
extinction]  a very similar
density of
formed stars, $\sim 3 \times 10^{-3}\rho_c$.

\section{Damped Lyman-$\alpha$ clouds as protospirals}

\begin{figure}
\vspace{6.0cm}
\includegraphics{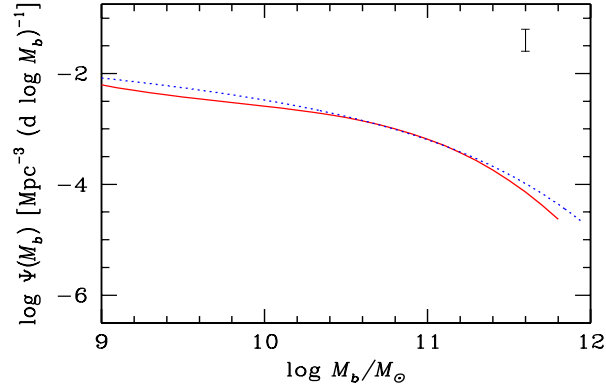}
\caption{Differential baryonic-mass number density of Ly$\alpha$
clouds (dotted line) and spirals (solid line). Also shown is the estimated
2$\sigma$ uncertainty on the DMF. 
}
\end{figure}

At high redshifts ($z \sim 2-3$), the Damped Ly$\alpha$ clouds,
i.e. 
neutral-hydrogen absorbers with column densities ${\cal
N}_{HI}> 2 \times
10^{20}$ atoms cm$^{-2}$, dominate the cosmological mass
density of the
gas in cosmic structures. The column density distribution
function,
$f({\cal N}_{HI})\, d{\cal N}_{HI}\, dx$\footnote {The absorption
distance
is $dx=(1+z)(1+2q_0z)^{-1/2}dz$, with $q_0$ the deceleration
parameter.}
(i.e., the number of absorbers per absorption distance interval
$dx$ and
column density interval $d{\cal N}_{HI}$, see Storrie-Lombardi et
al.
1996a), can be represented by:
$$
f({\cal N}_{HI}) ~=~ A ~\biggl({{\cal
N}_{HI} \over {\cal N}_*}\biggr)^{-1.5} e^{-\big({{\cal N}_{HI} \over
{\cal
N}_\star}-1\big)}\,,
\eqno(10)
$$ 
with ${\cal N}_*=10^{21.5}$ cm$^{-2}$ and $A=10^{23.8 \pm
0.4}$. The 
corresponding mean cosmological mass density is (e.g., Tytler
1987): 
$$
\Omega_{_{DLA}} ~=~ {H_0 \mu m_p \over c\rho_c} \int ^{{\cal 
N}_{HI}^{max}}_{{\cal 
N}_{HI}^{min}} {\cal N}_{HI}~ f({\cal N}_{HI},z)  ~d{\cal N}_{HI}
\eqno(11)
$$
(with log$\, {\cal N}_{HI}^{min}=20.3$, log$\, {\cal
N}_{HI}^{max}=21.8$,
$\mu=0.6$ the mean molecular weight, $m_p$ the proton mass,
and $c$ the
speed of light), and amounts to (Lanzetta et al. 1995;
Storrie-Lombardi et
al. 1996b): 
$$
\Omega_{_{DLA}} ~\sim~ (1\pm 0.35) \times
10^{-3}(\Omega_0+1) \,.
\eqno(12)
$$ 

Comparing eqs.(8) and (12) provides more precise evidence that
the baryon
content of spirals equals that of DLAs, $\omes \simeq
(0.9 -1.4) \Omega_{_{DLA}}$. [The earlier PS92 estimate was  
$\omes \simeq
(0.4 -1.2)
\,\Omega_{_{DLA}}$.] Indeed, the baryonic content of spirals just 
reaches
that of DLAs [the opposite claim (e.g., Storrie-Lombardi et
al.
1996b) is induced by overestimates of the spiral $M_b/L_B$
ratios]: this
coincidence is straightforwardly explained if DLAs are the
progenitors of
present-day spirals (Wolfe et al. 1986; Tytler 1987; Maloney
1992;
Lanzetta et al. 1995; Kauffmann 1996).  Furthermore, the
existence of a
narrow mass range at $\sim 10^{11}M_\odot$ for the
$\Omega_b$-dominating
objects may give  support to the single-population  model
which
identifies DLAs as discs rotating at one same high rotation
speed, 
$V_{rot} \sim 250$ km s$^{-1}$ (Prochaska \& Wolfe 1997,
1998). 

Let us investigate further the putative link between DLAs and
spirals by  comparing the cosmological distribution of their masses.
The scenario in which DLAs are progenitors of $z=0$ discs yields
a one-to-one 
link between a present-day spiral of mass $M_b$ and a high-$z$
DLA of mass 
$M_{Lya}$. With the assumption that spirals have retained most
of the baryon 
content of their DLA progenitors (assumption supported by the similar
cosmological 
densities of the two populations), we postulate:
$\psi^{\rm Ly\alpha} 
(M_{\rm Ly\alpha})=\psi^{\rm S} (M_{b})$. It is, then, remarkable
that the 
latter is fully consistent with the column density distribution of
eq.(10) if a 
suitable relationship exists between the central face-on column
densities
and the DLA masses. In fact, 
discs of the same mass having exponential surface density
profiles with same 
lengthscale will produce a range of observed column densities,
both because of 
having different inclinations and because of having a range of
impact parameters 
for different lines of sight. So, if we model DLAs as exponential
discs 
truncated at 5 lengthscales with vertical-to-radial lengthscale
ratio of $0.1-
0.3$, by averaging over all lines of sight and inclination angles
we get: 
${\cal N}_{HI}^0= {\cal N}_{HI} f^{-1}$, with $f \simeq 0.7-0.5$
(being $f 
\simeq {2 \over n^2} [1-(1+n)e^{-n}]\, {\rm ln}({\rm tg}\, {z_D
\over 
2nR_D})^{-2}$ with $z_D$ the disc vertical lengthscale). Then, 
let us assume ${\cal N}_{HI}^0/N_* = [M_{Ly\alpha}/ 
(aM_b^\oplus)]^\beta$: when $a=0.3$ and $\beta=0.45$, the
DMF and the DLA 
mass function essentially coincide (see Fig.6).

\section{Concluding remarks}

The baryonic mass function of disc systems is essentially a power law,
$\psi^{\rm
S}(M_b) d {\rm log} ~ M_b\propto M_b^{-1/2} d {\rm log} ~M_b$,
from $10^8 
M_\odot$ to $M_b^{\oplus} \sim 2 \times 10^{11} M_\odot$,
where a sudden 
cutoff occurs because of the lack of objects. In terms of their
baryonic 
content, spirals have a preferred mass scale $M_b^{\oplus}$,
around which 
most of the baryons with angular momentum are locked. 

The main feature of the luminosity function, an exponential
decline at
$L_B^*$, does not show up in the DMF. The (astrophysical)
features of the
luminosity function do not relate with the (cosmological) ones
emerging
from the baryonic mass function. 
         
The cosmological density of the luminous matter in spirals is a
negligible
fraction, $\sim 1/50$, of all the baryons synthesized in the
Big Bang,
and it is in good agreement with the cosmological density of
high-$z$ DLA
clouds.

In addition, we propose a continuity between the DLA clouds
and spiral 
galaxies: their (baryon) mass functions implying that, at $z=3$,
$\sim 10^{-4}$ 
objects Mpc$^{-3}$ with baryonic masses of $\sim 2 \times
10^{11} M_\odot$, 
total masses of $\sim 4 \times 10^{12}M_\odot$, and dynamical
times of $\sim$1 
Gyr, are the protagonists of structure in the Universe.

\vskip 1.2truecm

\centerline{\bf APPENDIX}

\vskip 0.8cm

In Table 1 we report the Schechter parameters of the LFs used in the
paper. 

\begin{figure}
\vspace{7.9cm}
\includegraphics{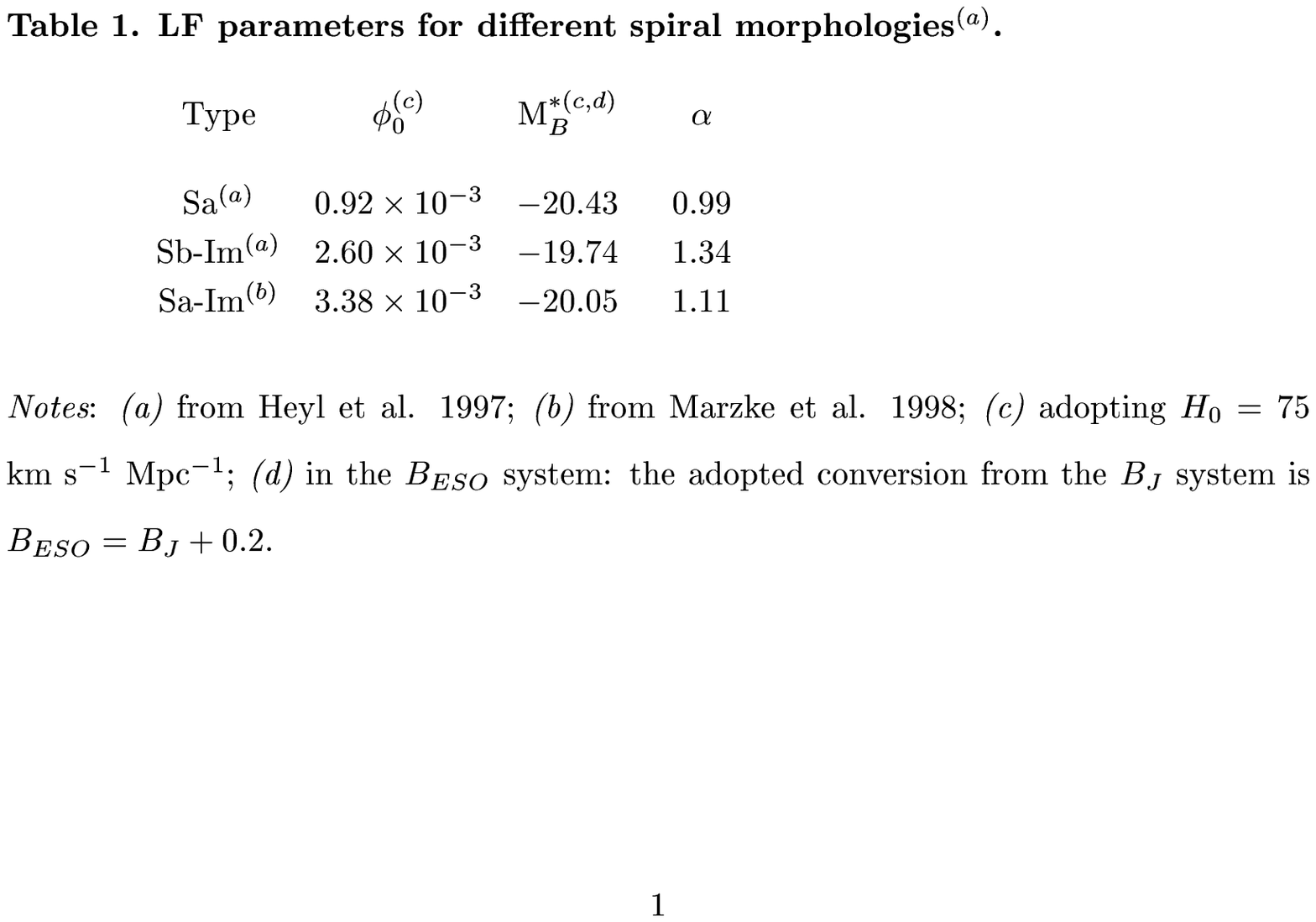}
\label{fig:TABLE}
\end{figure}

In Fig.7 we show the DMF constructed using the $M_b(L_B)$ relation 
described by 
eqs.(3)-(5) and the spiral LF of Marzke et al. (1998).

\begin{figure}
\vspace{7.0cm}
\includegraphics{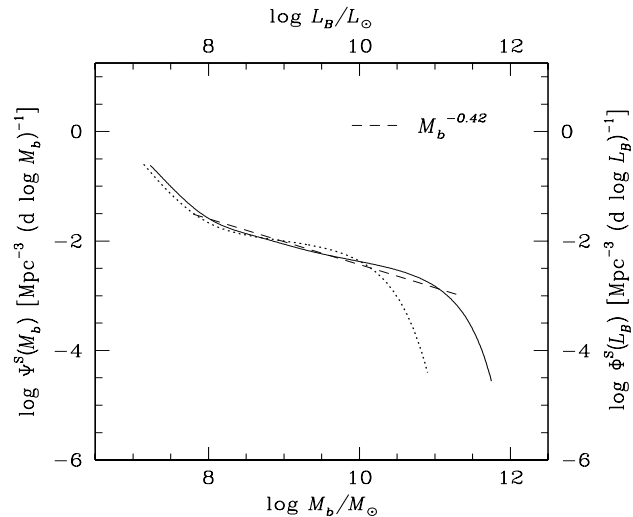}
\caption{
The spiral galaxy luminosity function of Marzke et al. 1998 (dotted line) 
and the corresponding baryonic mass function, from the $M_b(L_B)$
relation in eqs.(3)-(5) (solid line). 
}
\end{figure}

\newpage
\section { References}
\vglue 0.2truecm

\ref{Bertola, F., Pizzella, A., Persic, M., \& Salucci, P. 1993, ApJ, 416, L45}
\ref{Bertola, F., et al. 1998, ApJ, 509, L93}
\ref{Bland-Hawthorn, J., Freeman, K.C., \& Quinn, P.J. 1997, ApJ, 490, 143}
\ref{Corbelli, E., \& Schneider, S.E. 1997, ApJ, 479, 244}
\ref{Corsini, E.M., et al. 1999, A\&A, 342, 671}
\ref{de Blok, W.J.G., McGaugh, S.S., \& van der Hulst, J.M.
	1996, MNRAS, 283, 18}
\ref{Efstathiou, G., Ellis, R.S., \& Peterson, B.A. 1988, MNRAS,
	232, 431}
\ref{Ellis, R.S., Colless, M., Broadhurst, T., Heyl, J., \& 
     Glazebrook, K. 1996, MNRAS, 280, 235}
\ref{Heyl, J., Colless, M., Ellis, R.S., \& Broadhurst, T. 1997, 
     MNRAS, 285, 613} 
\ref{Hoffman, G.L., Salpeter, E.E., Farhat, B., Roos, T., Williams,
     H., \& Helou, G. 1996, ApJS, 105, 269}
\ref{Hogg, D.W., Cohen, J.G., Blandford, R., \& Pahre, M.A.
     1998, ApJ, 504, 622}
\ref{Honma, M., \& Sofue, Y. 1997, PASJ, 49, 453}
\ref{Jablonka, P., \& Arimoto, N. 1992, A\&A, 255, 63}
\ref{Kauffmann, G. 1996, MNRAS, 281, 475}
\ref{Lanzetta, K.M., Wolfe, A.M., \& Turnshek, D.A. 1995, ApJ,
	440, 445}
\ref{Lin, H., et al. 1996, ApJ, 464, 60}
\ref{Loveday, J. 1998, in Proc. XVIII Moriond Astrophysics Meetings 
	"Dwarf Galaxies and Cosmology", ed. Thuan et al. (Editions 
	Fronti\`eres), in press (astro-ph/9805255)}
\ref{Loveday, J., Peterson, B.A., Efstathiou, G., \& Maddox, S.J.
	1992, ApJ, 390, 338}
\ref{Maloney, P. 1992, ApJ, 398, L89}
\ref{Marzke, R.O., Huchra, J.P., \& Geller, M.J. 1994, ApJ, 428, 43} 
\ref{Marzke, R.O., et al. 1998, ApJ, 503, 617}
\ref{Oliva, E., Origlia, L., Kotilainen, J.K, and Moorwood, A.F.M.
	1995, A\&A, 301, 55}
\ref{Persic, M., \& Salucci, P. 1990a, MNRAS, 245, 577}
\ref{Persic, M., \& Salucci, P. 1990b, MNRAS, 247, 349}
\ref{Persic, M., \& Salucci, P. 1992, MNRAS, 258, 14P (PS92)}
\ref{Persic, M., Salucci, P., \& Stel, F. 1996, MNRAS, 281, 27}
\ref{Prochaska, J.X., \& Wolfe, A.M. 1997, ApJ, 487, 73}
\ref{Prochaska, J.X., \& Wolfe, A.M. 1998, ApJ, 507, 113}
\ref{Quillen, A.C., \& Frogel, J.A. 1997, ApJ, 487, 603}
\ref{Radcliffe, A., Shanks, T., Parker, Q.A., \& Fong, R. 1998, 
     MNRAS, 293, 197}
\ref{Rhee, M.-H. 1996, PhD thesis, University of Groningen}
\ref{Rees, M.J., \& Ostriker, J.P. 1977, MNRAS, 179, 541}
\ref{Salpeter, E.E., \& Hoffman, G.L. 1996, ApJ, 465, 595}
\ref{Salucci, P., Ashman, K.M., \& Persic, M. 1991, ApJ, 379, 89}
\ref{Salucci, P., Frenk, C.S., \& Persic, M. 1993, MNRAS, 262, 392}
\ref{Salucci, P., \& Persic, M. 1997, in "Dark and Visible Matter
	in Galaxies", ed. M.Persic \& P.Salucci, ASP Conference Series
	(San Francisco: Astronomical Society of the Pacific), 117, 1 }
\ref{Salucci, P., et al. 1999, in preparation}
\ref{Schechter, P.L. 1976, ApJ, 203, 297 }
\ref{Schechter, P.L., \& Dressler, A. 1987, AJ, 94, 563}
\ref{Silva, L., Granato, G.L., Bressan, A., \& Danese, L. 1998, 
	ApJ, 509, 103}
\ref{Sprayberry, D., Impey, C.D., Irwin, M.J., \& Bothun, G.D.
1997, 
     ApJ, 482, 104}
\ref{Storrie-Lombardi, L.J., Irwin, M.J., \& McMahon, R.G.
	1996a, MNRAS, 282, 1330}
\ref{Storrie-Lombardi, L.J., McMahon, R.G., \& Irwin, M.J.
	1996b, MNRAS, 283, L79}
\ref{Thoul, A.A. \& Weinberg, D.H. 1996, ApJ, 465, 608 }
\ref{Tytler, D. 1987, ApJ, 321, 69}
\ref{Willick, J.A. 1998, astro-ph/9809160}
\ref{Wolfe, A.M., Turnshek, D.A., \& Smith, H.E., \& Cohen, R.D.
	1986, ApJS, 61, 249}
\ref{Zucca, E., et al. 1997, A\&A, 326, 477}
\ref{Zwaan, M.A., Briggs, F.H., Sprayberry, D., \& Sorar, E. 1997, 
     ApJ, 490, 173 }

\end{document}